\documentclass[11pt]{article}
\usepackage{graphics}
\usepackage{overpic}
\usepackage{subfigure}
\usepackage{graphicx}
\usepackage{color}
\usepackage{bm}
\usepackage{amsfonts}
\usepackage{amsmath}
\usepackage{mathrsfs}
\usepackage[square,numbers,sort&compress]{natbib}
\usepackage{caption}
\usepackage{float}
\usepackage[dvipdfm]{hyperref}

\allowdisplaybreaks

\newcommand{\me}{\mathrm{e}}
\newcommand{\mi}{\mathrm{i}}

\def\lam{\lambda}
\def\mb{\mathbf}
\def\ra{\rightarrow}

\def\lim{\mathrm{lim}}
\def\sech{\mathrm{sech}}

\def\Im{\mathrm{Im}}
\def\sgn{\mathrm{sgn}}
\def\arg{\mathrm{Arg}}

\def\max{\mathrm{max}}
\def\NL{\mathrm{NL}}

\def\eref#1{(\ref{#1})}
\begin{document}
\date{}

\title{Anti-dark and Mexican-hat solitons in the Sasa-Satsuma equation on the continuous wave background}

\author{Tao Xu$^{1}$, Min Li$^{2}$ and Lu Li$^{3}$
 \\{\em 1. College of Science,  China University of Petroleum, Beijing 102249,
China,}
\\{\em E-mail: xutao@cup.edu.cn.}
\\{\em 2. Department of Mathematics and Physics, North China Electric}
\\{\em Power University, Beijing 102206, China.}
\\{\em 3. Institute of Theoretical Physics, Shanxi University, Taiyuan, Shanxi
030006, China.}} \maketitle

\begin{abstract}
In this letter, via the Darboux transformation method we construct
new analytic soliton solutions for the Sasa-Satsuma equation which
describes the femtosecond  pulses propagation in a monomode fiber.
We reveal that two different types of femtosecond solitons, i.e.,
the anti-dark (AD) and Mexican-hat (MH) solitons, can form on a
continuous wave (CW) background, and numerically study their
stability under small initial perturbations.  Different from the
common bright and dark solitons,  the AD and MH solitons can exhibit
both the resonant and elastic interactions, as well as various
partially/completely inelastic interactions which are composed  of
such two fundamental interactions.  In addition, we find that the
energy exchange between some interacting soliton and the CW
background may lead to one AD soliton changing into an MH one, or
one MH soliton into an AD one.

$\;$


\end{abstract}

$\;$

\emph{Introduction.} --- As localized wave packets formed by the
balance between the group-velocity dispersion and self-phase
modulation~\cite{HT1973}, solitons in optical fibers have drawn
considerable attention because of their robust nature and potential
application in all-optical, long-distance
communications~\cite{NonlOptics}. In the picosecond regime, the
model governing the propagation of optical solitons  in a
single-mode fibre is the celebrated nonlinear Schr\"{o}dinger
equation (NLSE)~\cite{HT1973}. However, one has to take into account
some higher-order linear and nonlinear effects for the ultrashort
pulses propagating in high-bit-rate transmission
systems~\cite{NonlOptics}.
The governing model for the femtosecond  pulse propagation is the
following higher-order NLSE~\cite{Kodama1}:
\begin{eqnarray}
\mi\,{u}_{z} + \frac{\sigma}{2}\,{u}_{tt}+ |u|^{2}u = -
\mi\,\varepsilon\! \left[\sigma_{1}\,u_{ttt} +
\sigma_{2}(|u|^{2}u)_t+\sigma_{3}\,u(|u|^{2})_{t}\right] ,
\label{HNLS}
\end{eqnarray}
where $\sigma = \pm 1$, $\varepsilon $ is a real small parameter, $
\sigma_{1} $, $ \sigma_{2} $ and $ \sigma_{3} $  represent the
third-order dispersion (TOD), self-steepening (SS, also known as
Kerr dispersion) and stimulated Raman scattering (SRS) effects,
respectively~\cite{NonlOptics,Kodama1}.

With $\sigma_1 \neq 0$, Eq.~\eref{HNLS}  has two important
integrable versions: 
(i) the Hirota equation (HE)~\cite{Hirota}, $ \sigma_{1}:
\sigma_{2}: \sigma_{2}+ \sigma_{3} =1: 6\,\sigma: 0 $; (ii) the
Sasa-Satuma equation (SSE)~\cite{SS}, $ \sigma_{1}: \sigma_{2}:
\sigma_{2} + \sigma_{3} =1: 6\,\sigma :3\,\sigma $.  The SSE usually
takes the form~\cite{SS}
\begin{align}
\mi\,{u}_{z} + \frac{\sigma}{2}\,{u}_{tt}+ |u|^{2}u +
\mi\,\varepsilon\! \left[u_{ttt} + 6\,\sigma (|u|^{2}u)_{t} -
3\,\sigma u(|u|^{2})_{t}\right] = 0. \label{SS}
\end{align}
Although there is a fixed relation among the higher-order terms, the
SSE is thought to be more fundamental than the HE for applications
in optical fibers because the former contains the SRS
term~\cite{NonlOptics,Kodama1}. Up to now, many integrable
properties of Eq.~\eref{SS} have been detailed, like the inverse
scattering transform scheme~\cite{SS,Mihalache}, bilinear
representation~\cite{Gilson}, Painlev\'{e} property~\cite{Painleve},
conservation laws~\cite{Kim}, nonlocal symmetries~\cite{Sergyeyev},
squared eigenfunctions~\cite{Yang}, B\"{a}cklund
transformation~\cite{Backlund} and Darboux transformation
(DT)~\cite{SSDT1,SSDT2}.

The presence of the SRS term enriches the solitonic behavior in
Eq.~\eref{SS}~\cite{SS,Mihalache,Gilson,SSDT2,LZH,Gedalin,Jiang,Ohta}.
Under the vanishing boundary condition (VBC), Eq.~\eref{SS} with
$\sigma=1$ possesses the common single-hump
soliton~\cite{SS,Gedalin}, the double-hump soliton behaving like two
in-phase solitons with a fixed separation~\cite{SS,Mihalache}, and
the multi-hump breather with the periodically-oscillating
structure~\cite{Mihalache,Gilson}. Under the non-vanishing boundary
condition,  Eq.~\eref{SS} with $\sigma=-1$ admits the common dark
soliton, and the double-hole dark soliton displaying two symmetric
dips with a fixed separation~\cite{Jiang,Ohta}; while Eq.~\eref{SS}
with $\sigma=1$ has the bright-like soliton which is linearly
combined of a dark one and a bright one~\cite{LZH}.
On the other hand, the  soliton interaction behavior underlying in
the SSE is far more abundant and complicated than that in the NLSE.
Even with the VBC, the shape-changing interactions between soliton
and breather have recently been found in Eq.~\eref{SS} with
$\sigma=1$~\cite{SSDT2}.

In this letter, we are trying to reveal some novel solitonic
phenomena on a  continuous wave (CW) background for Eq.~\eref{SS}
with $\sigma=1$. Via the DT technique~\cite{SSDT2}, we obtain three
families of single-soliton solutions which can display two
completely-different profiles.  The first type is the anti-dark (AD)
soliton having the form of a bright soliton on a CW background,
i.e., it looks like a dark soliton with reverse sign
amplitude~\cite{Kivshar}. The second type takes the Mexican-hat (MH)
shape, that is, one high hump carries two small dips which have a
symmetrical distribution with respect to the hump, hence such new
type of soliton is called the MH soliton. More importantly, we find
that the femtosecond AD and MH solitons  admit  the resonant
interaction, elastic interaction, as well as various
partially/completely inelastic interactions which consist of the
fundamental resonant and elastic interaction structures.
To our knowledge, it is the first time that the coexistence of
elastic and resonant soliton interactions has been found in the
NLSE-type models.
Physically, the resonant interaction of optical waves excited from a
CW background can be used to realize a second-harmonic generation in
the centro-symmetry optical fiber~\cite{Cui}.

\emph{N-th iterated soliton solutions via the Darboux
transformation.} ---  With the simple CW solution $u= \rho\,\me^{\mi
(\frac{t}{6 \varepsilon }-\frac{z}{108 \varepsilon^2}+\phi)}$
($\rho>0$ and $\phi$ are both real constants) as a seed,  we employ
the DT-iterated algorithm presented in Ref.~\cite{SSDT2} to obtain
the N-th iterated solution in the form
\begin{align}
& u_N = 
\me^{\mi (\frac{t}{6 \varepsilon }-\frac{z}{108 \varepsilon^2})}
\left(\rho\,\me^{\mi \phi} -
2\,\frac{\tau_{N+1,N-1,N}}{\tau_{N,N,N}}\right),
\label{potentialTran2}
\end{align}
with
\begin{align}
\tau_{J,K,L} = \begin{vmatrix}
\mb{F}_{N\times J}  & -\mb{G}_{N\times K}  &  -\mb{H}_{N\times L}   \\
\mb{F}^*_{N\times J} & -\mb{H}^*_{N\times K}  & -\mb{G}^*_{N\times L}  \\
\mb{G}^*_{N\times J} & \mb{F}^*_{N\times K}  & \mb{0}   \\
\mb{H}_{N\times J}   & \mb{F}_{N\times K} & \mb{0}  \\
\mb{H}^*_{N\times J} & \mb{0} & \mb{F}^*_{N\times L}\\
\mb{G}_{N\times J}   & \mb{0}  &  \mb{F}_{N\times L}
\end{vmatrix} ,  \label{tau}
\end{align}
where $J+K+L=6\,N $, the block matrices $
 \mb{F}_{N\times J} = \big(\lam_k^{m-1}f_k\big)_{\substack{1\leq k \leq N,\\ 1\leq m \leq J}}$,
 $\mb{G}_{N\times K} = \big[(-\lam_k)^{m-1}g_k\big]_{\substack{1\leq k \leq
 N,\\ 1\leq m \leq K}}$, $\mb{H}_{N\times L} =
\big[(-\lam_k)^{m-1}h_k\big]_{\substack{1\leq k \leq N, \\ 1\leq m
\leq L}}$, and the functions $f_k$, $g_k$, $h_k$  $(1\leq k \leq N)$
are given as
\begin{eqnarray} \left\{
\begin{aligned}
& f_k =\me^{\frac{\mi \phi }{2}} \left(\alpha_k \me^{\theta_k (t,z)} + \beta_k \me^{-\theta_k (t,z)} \right),  \\
& g_k = -\me^{-\frac{\mi\,\phi }{2}} \left(\frac{ \rho
\alpha_k}{\chi_k^+} \me^{\theta_k (t,z)} +\frac{ \rho
\beta_k}{\chi_k^-}\me^{-\theta_k (t,z)} -\gamma_k \,
\me^{-\omega_k(t,z)}  \right),  \\
& h_k = -\me^{\frac{3 \,\mi\,\phi }{2}} \left(\frac{ \rho
\alpha_k}{\chi_k^+}\me^{\theta_k (t,z)} +\frac{ \rho
\beta_k}{\chi_k^-}\me^{-\theta_k (t,z)} + \gamma_k
\me^{-\omega_k(t,z)} \right),   \label{LPsolutions}
\end{aligned}\right.
\end{eqnarray}
 with $\theta_k (t,z)  = \chi_k \left[t-\frac{z}{12\,
\varepsilon}-4 \left(\lam_k ^2+\rho ^2\right) \varepsilon\,z
\right]$, $\omega_k(t,z) =
\lam_k\left(t-\frac{z}{12\,\varepsilon}-4\,\lam_k^2 \varepsilon\,z
\right)$, $\chi_k  = \sqrt{\lam_k^2-2 \rho ^2}$, $\chi_k^{\pm} =
\lam_k \pm \sqrt{\lam_k^2-2 \rho ^2}$, $ \alpha_k $, $ \beta_k $ and
$\gamma_k$ being nonzero complex constants.


In order to obtain the solitonic structure from
solution~\eref{potentialTran2}, we require all $\chi_k $'s ($1\leq k
\leq N$) be  real numbers, that is, $\Im(\lam_k)=0 $ and $|\lam_k|>
\sqrt{2}\,\rho$.
For convenience of our analysis, we introduce the notations
$\mu^{(1)}_k = \alpha_k^*\beta_k -\alpha_k \beta_k^*$, $ \mu^{(2)}_k
= \alpha^*_k \gamma_k +\alpha_k \gamma^*_k$ and $\mu^{(3)}_k  =
\beta^*_k \gamma_k + \beta_k \gamma^*_k$ ($k=1,2)$, and use ``$(n,
m)$'' to represent the soliton interaction  with $n$ asymptotic
solitons as $z\ra -\infty$ and $m$ ones as $z\ra \infty$.

\begin{figure}[H]
 \centering
\subfigure[]{\label{Fig1}
\includegraphics[scale=0.3]{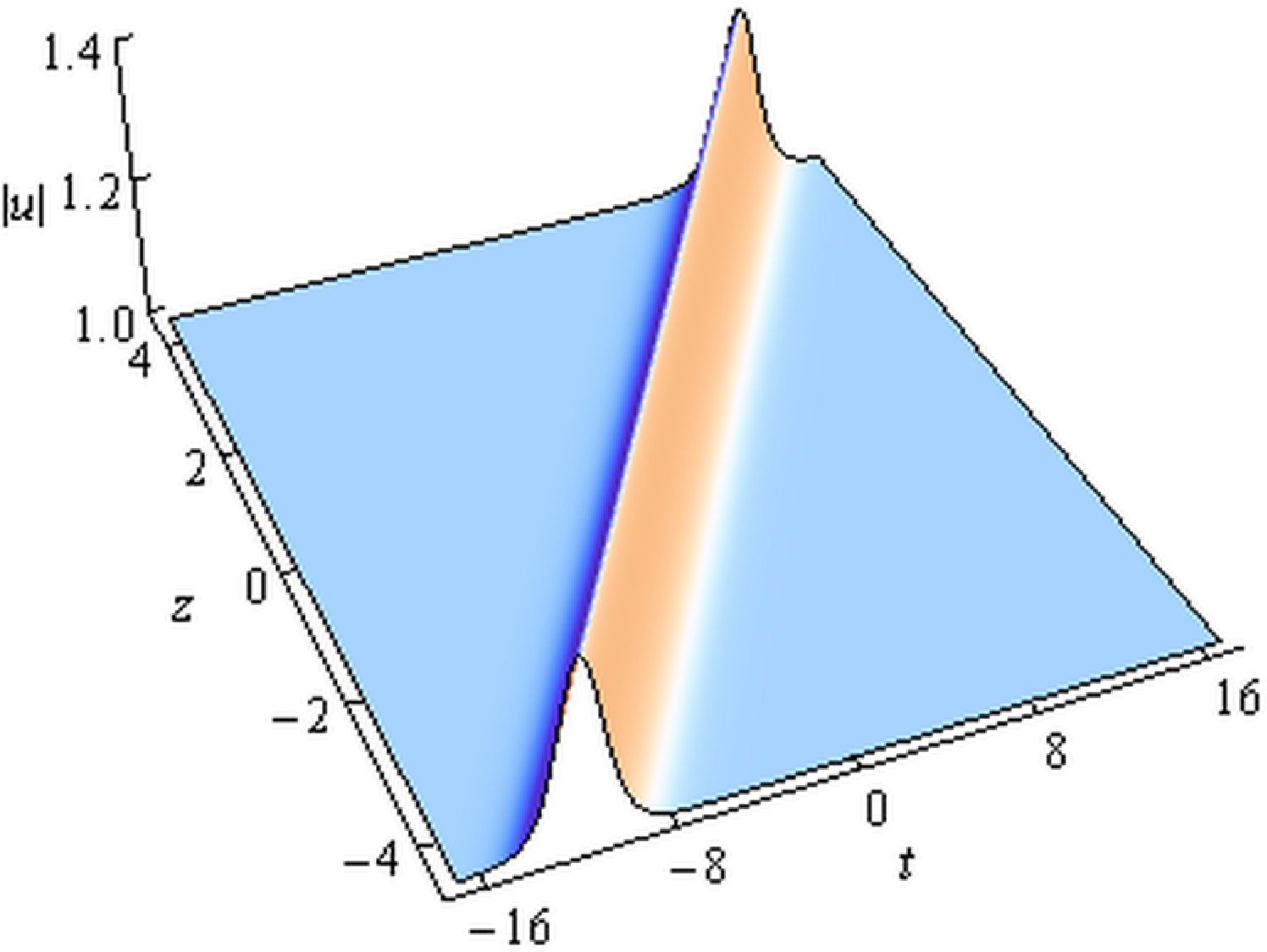}}\hfill
\subfigure[]{ \label{Fig2}
 \includegraphics[scale=0.3]{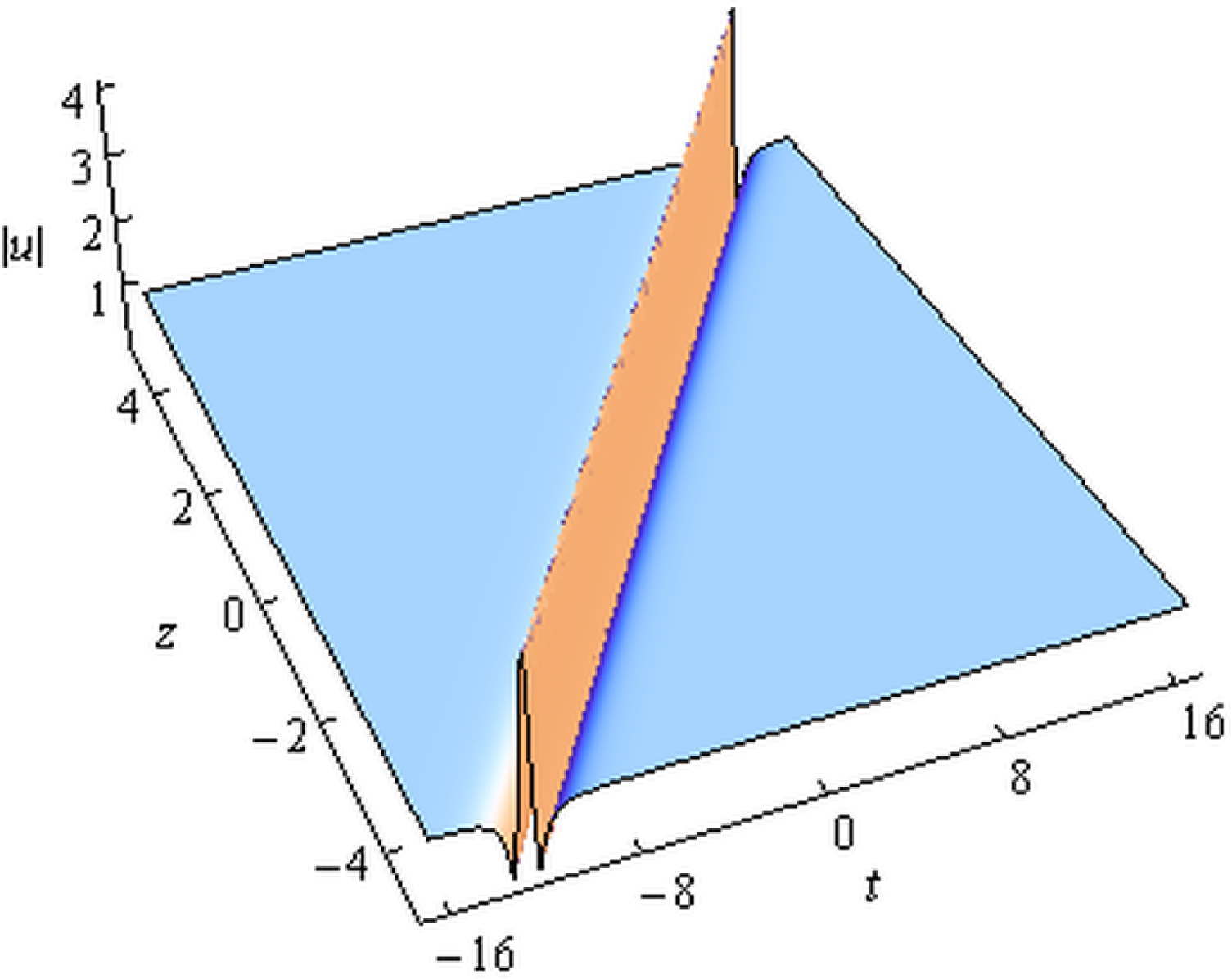}}\hfill
\subfigure[]{ \label{Fig2b}
\includegraphics[scale=0.3]{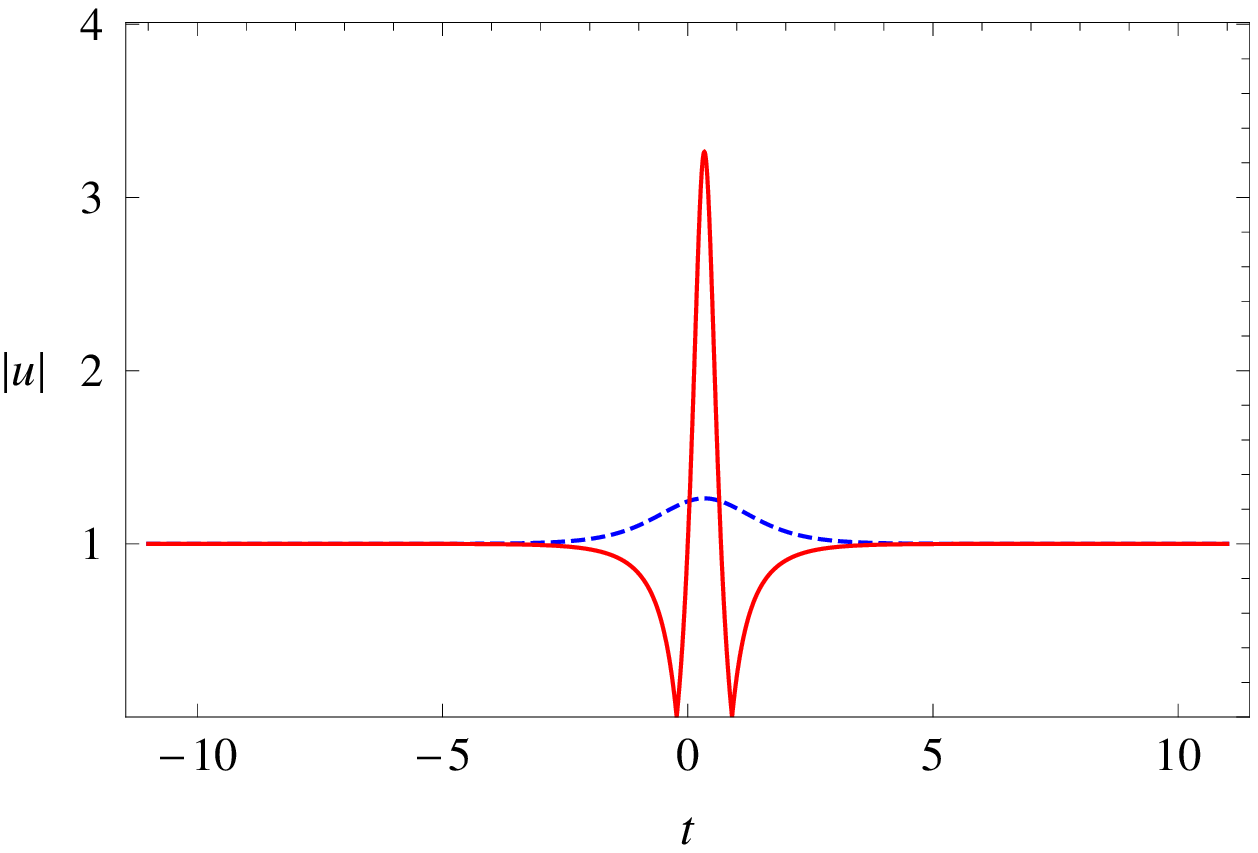}}
\caption{  Evolution of the (a) AD soliton  and (b) MH soliton
plotted via solution~\eref{Solution1}, where
$|\alpha_1|=|\beta_1|=1$, $\lam_1=1.6$, $\rho=1$, $\phi=0$,
$\varepsilon=0.15$, $\phi^{(1)}_1=0$ for (a) and $\phi^{(1)}_1=\pi$
for (b).  (c) Transverse plots of AD (blue dotted line) and MH (red
solid line) solitons at $z=0$.}
\end{figure}

\emph{Anti-dark and Mexican-hat solitons.} ---  For
solution~\eref{potentialTran2} with $N=1$, we can obtain three
families of single-soliton solutions  under the reducible cases
$\mu^{(i)}_1=0$ ($1 \leq i\leq 3$). If $\mu^{(1)}_1=0 $ (i.e.,
$\beta_1 \alpha^*_1 - \alpha_1 \beta^*_1 =0 $), the solution can be
written as
\begin{align}
u^{(1)}_1= & \rho\,\me^{\mi (\frac{t}{6 \varepsilon }-\frac{z}{108
\varepsilon^2}+\phi +\pi)} + \frac{\sqrt{2}\,\chi^2_1\,\me^{\mi
(\frac{t}{6 \varepsilon } -\frac{z}{108 \varepsilon^2}+\phi + \pi)}
}{\sqrt{2}\, \rho  +|\lam_1| \,\me^{\mi \,\phi^{(1)}_1} \cosh
\big(\Theta^{(1)}_1+\delta^{(1)}_1\big)}, \label{Solution1}
\end{align}
where $\Theta^{(1)}_1 = 2\, \theta_1 (t,z)$, $\delta^{(1)}_1 =
\frac{1}{2}\, \ln \left(\frac{|\alpha_1|^2
\chi_1^-}{|\beta_1|^2\chi_1^+}\right)$, $\phi^{(1)}_1=
\arg(\alpha_1)-\arg(\beta_1) =0\, \text{or}\, \pi $. In this
solution, the first part $\rho\,\me^{\mi (\frac{t}{6 \varepsilon
}-\frac{z}{108 \varepsilon^2}+\phi +\pi)}$ is a CW solution of
Eq.~\eref{SS},
while the second part describes a soliton embedded in the CW
background (Note that the denominator has no singularity because
$|\lam_1|> \sqrt{2}\,\rho$).

The parameter $\phi^{(1)}_1=0 $ implies that the embedded solution
has the same phase as that of the CW solution. In this case,
$u^{(1)}_1$ represents an AD soliton which displays the bright
soliton profile on the CW pedestal [see Figs.~\ref{Fig1}
and~\ref{Fig2b}]. The soliton velocity and width are, respectively,
given by $v= 4 \varepsilon \left(\rho ^2+\lam_1^2\right)+\frac{1}{12
\varepsilon }$ and $w=\frac{1}{2 \,\chi_1}$, and $|u^{(1)}_1|$
reaches the maximum $|u^{(1)}_1|_{\max} = \frac{\rho \left|\lam
_1\right|+\sqrt{2} \left(\lam_1^2-\rho ^2\right)}{\left(\left|\lam_1
\right| + \sqrt{2} \rho \right)} $ when   $\Theta^{(1)}_1 =0 $. If
$\phi^{(1)}_1=\pi $, the embedded solution and CW solution have the
same phases in the inner region $ \frac{\sqrt{2} \left(\rho^2 +
\chi_1^2\right) - \chi_1 \sqrt{2 \lam_1^2-\rho ^2} }{\rho
\left|\lam_1\right|} \leq \Theta^{(1)}_1 + \delta^{(1)}_1 \leq
\frac{\sqrt{2} \left(\rho^2 + \chi_1^2\right) + \chi_1 \sqrt{2
\lam_1^2-\rho ^2}}{\rho \left|\lam_1\right|}$, but their phases are
opposite in the outer region  $ \Theta^{(1)}_1+\delta^{(1)}_1 <
\frac{\sqrt{2} \left(\rho^2 + \chi_1^2\right) - \chi_1
\sqrt{2\lam_1^2-\rho ^2}}{\rho \left|\lam_1\right|} $ or $
\Theta^{(1)}_1 +\delta^{(1)}_1
> \frac{\sqrt{2} \left(\rho^2 + \chi_1^2\right) + \chi_1 \sqrt{2
\lam_1^2-\rho ^2}}{\rho\left|\lam_1\right|}$. Hence, the modulus of
$u^{(1)}_1$ exhibits that one high hump is symmetrically accompanied
with two small dips beneath the CW background, which looks like the
MH shape [see Figs.~\ref{Fig2} and~\ref{Fig2b}]. The velocity and
width of the MH soliton are the same as those of the AD one, but its
maximum amplitude drastically increases to $ |u^{(1)}_1|_{\max} =
\frac{\sqrt{2}\left(\rho^2+\chi_1^2\right)-\rho
\left|\lam_1\right|}{\left|\lambda _1\right|-\sqrt{2} \rho} $ at the
center of the hump, and drops to zero at the centers of two dips.
The generation of the MH soliton could be explained as that the
phase oppositeness makes  some energy be transferred  from the CW
background to the embedded solution, and further leads to rising of
one hump and  sinking of two dips.

For the reducible cases $\mu^{(2)}_1=0$ and $\mu^{(3)}_1=0 $, we can
obtain the other two families of single-soliton solutions as
follows:
\begin{align}
 u^{(2)}_1 = & \me^{\mi (\frac{t}{6 \varepsilon }-\frac{z}{108
\varepsilon^2}+\phi)} \Big[ \rho\, \tanh \big(\Theta^{(2)}_1 +
\delta^{(2)}_1 \big) + \me^{\mi \,\phi^{(2)}_1} \sqrt{\lam_1
\chi_1^+}\, \sech \big(\Theta^{(2)}_1 + \delta^{(2)}_1 \big)
\Big], \label{Solution2}\\
u^{(3)}_1=&  \me^{\mi (\frac{t}{6 \varepsilon }-\frac{z}{108
\varepsilon^2}+\phi + \pi )} \Big[\rho \tanh \big(\Theta^{(3)}_1 +
\delta^{(3)}_1 \big)  + \me^{\mi\,\phi^{(3)}_1}  \sqrt{\lambda _1
 \chi_1^-}\,\sech \big(\Theta^{(3)}_1 +
\delta^{(3)}_1 \big) \Big], \label{Solution3}
\end{align}
where $\Theta^{(2)}_1 = \theta_1 (t,z)+\omega_1(t,z)$,
$\Theta^{(3)}_1= \theta_1 (t,z)-\omega_1(t,z)$, $ \delta^{(2)}_1 =
\frac{1}{2} \, \ln \left(\frac{|\alpha_1|^2 \chi_1^2}{|\gamma_1|^2
\lam_1\chi_1^+}\right)$, $\delta^{(3)}_1 = \frac{1}{2}\,\ln \left(
\frac{|\gamma_1|^2 \lam_1\chi_1^- }{|\beta_1|^2 \chi_1^2}\right)$,
$\phi^{(2)}_1= \arg(\alpha_1)-\arg(\gamma_1)= \pm \frac{\pi}{2} $
and $\phi^{(3)}_1= \arg(\beta_1)-\arg(\gamma_1) = \pm \frac{\pi}{2}
$.  Because $|\lam_1|> \sqrt{2}\,\rho$, either $u^{(2)}_1$ or
$u^{(3)}_1$ displays only the AD soliton profile, which is similar
to the case $\phi^{(1)}_1=0 $ in solution~\eref{Solution1}.
Solutions~\eref{Solution2} and~\eref{Solution3} are also called the
combined solitary wave solutions~\cite{LZH}. The combined dark and
bright solitons have a constant phase difference $\frac{\pi}{2} $ or
$-\frac{\pi}{2}$. Such phase difference causes a nonlinear phase
shift, for example, the nonlinear phase shift in
solution~\eref{Solution2}  can be given as
\begin{align}
\phi^{(2)}_{\NL}(t,z) =\arctan \bigg[\sin \phi^{(2)}_1 \cdot
\frac{\sqrt{\lam_1 \chi_1^+}\,\sech \big( \Theta^{(2)}_1 +
\delta^{(2)}_1\big)}{\rho \tanh \big( \Theta^{(2)}_1 +
\delta^{(2)}_1\big)}  \bigg].
\end{align}

\begin{figure}[H]
 \centering
\subfigure[]{\label{FigIIa}
\includegraphics[scale=0.25]{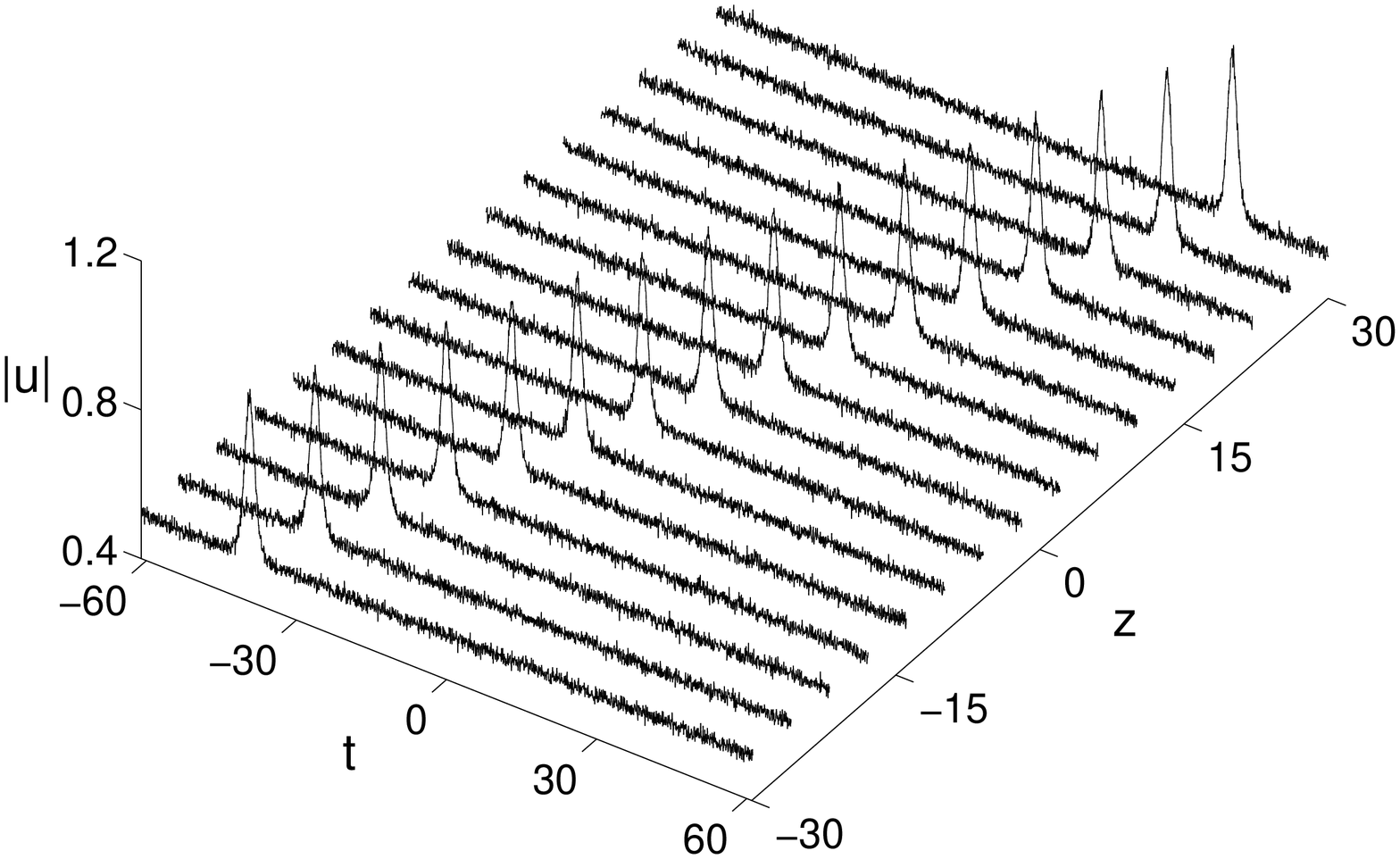}}\hfill
\subfigure[]{\label{FigIIb}
 \includegraphics[scale=0.27]{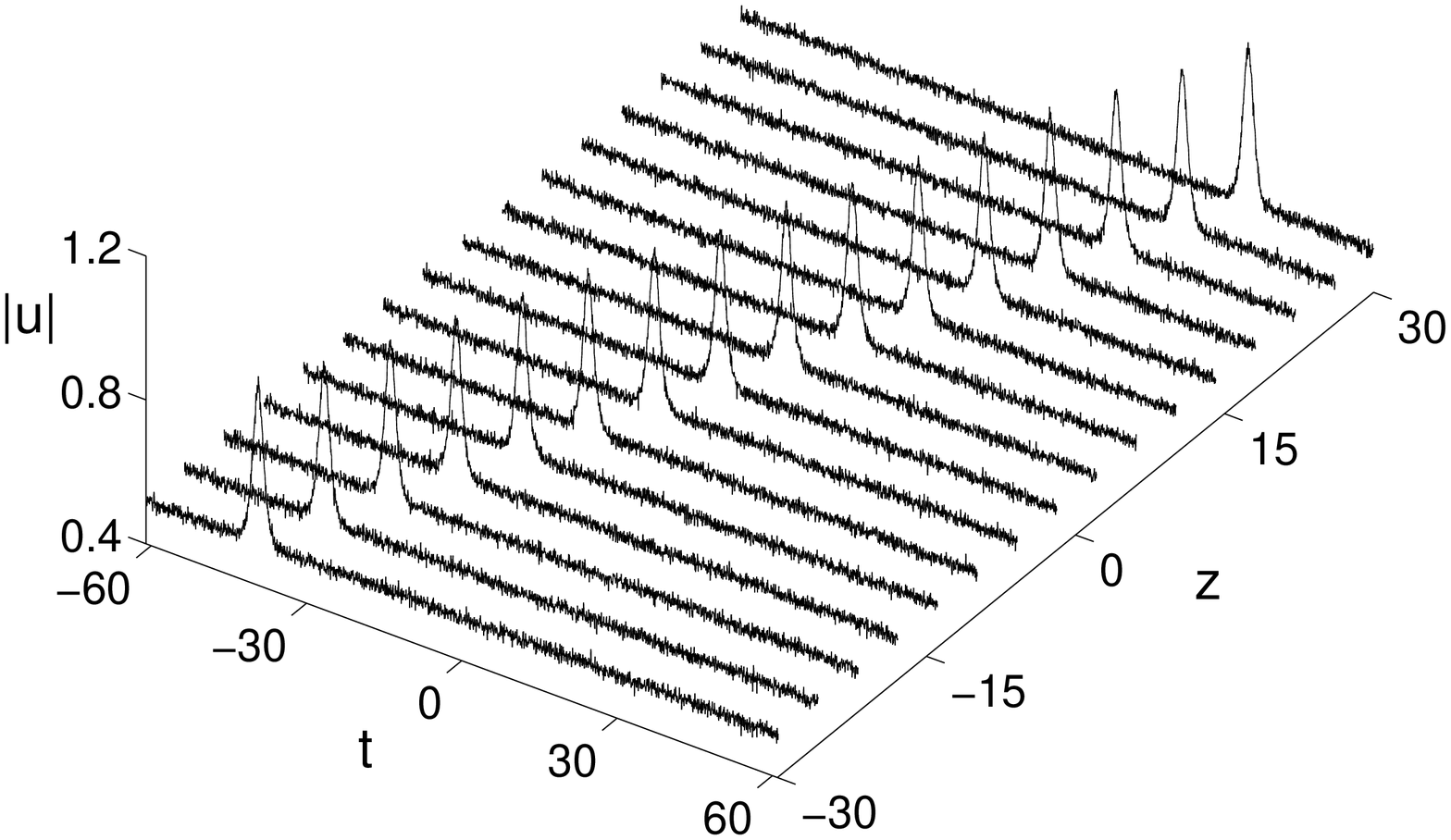}}
\caption{Numerical evolution of the AD soliton under the
perturbation of a white noise with the maximal value $0.08$. The
initial pulse corresponds to solution~\eref{Solution1} at $z=0$ with
the parameters as $\alpha_1=\beta_1 =\lam_1=1$, $\rho=0.5$,
$\varepsilon=0.13$, (a) $\sigma_{1}: \sigma_{2}: \sigma_{2} +
\sigma_{3} =1:6:3$, (b) $\sigma_{1}: \sigma_{2}: \sigma_{2} +
\sigma_{3} =1:5.8:2.5$.}
\end{figure}

The stability of solitons is a crucial issue for their applications
in optical communication lines~\cite{NonlOptics}. It has been shown
in Ref.~\cite{LZH} that the solitons described by
solutions~\eref{Solution2} and~\eref{Solution3} enjoy a good
stability against finite-amplitude initial perturbations.  Here, the
numerical simulation is performed to examine the stability of
solution~\eref{Solution1}  by the split-step Fourier
method~\cite{NonlOptics}.
Fig.~\ref{FigIIa} shows that with the presence of a white noise, the AD soliton can propagate stably for
$60$ dispersion lengths along the fiber.  Also, we numerically
simulate the evolution of the AD soliton 
when the TOD, SS and SRS terms do not obey the
fixed relation in Eq.~\eref{SS}.
Fig.~\ref{FigIIb} illustrates that the AD soliton still keeps its
stable shape after propagating $60$ dispersion lengths  when $
\sigma_{1}: \sigma_{2}: \sigma_{2} + \sigma_{3}= 1: 5.8: 2.5$ and
the initial pulse is perturbed by a white noise. However,  there is
some radiation on the background of the MH soliton during the
propagation if a white noise is added in the initial pulse. We note
that the practical optics telecommunication system is usually
dissipative because of the  fiber loss/gain~\cite{Liu2}. With the
inclusion of a linear loss/gain term into Eq.~\eref{SS}, our
numerical experiments show that both the AD and MH solitons together
with the CW background decay/grow exponentially with the evolution
of $z$. Thus, the  balance between the energy input and output also
plays an important role in maintaining a long-lived optical
soliton~\cite{Liu3}.

\begin{figure}[H]
 \centering
\subfigure[]{\label{Fig3}
\includegraphics[scale=0.45]{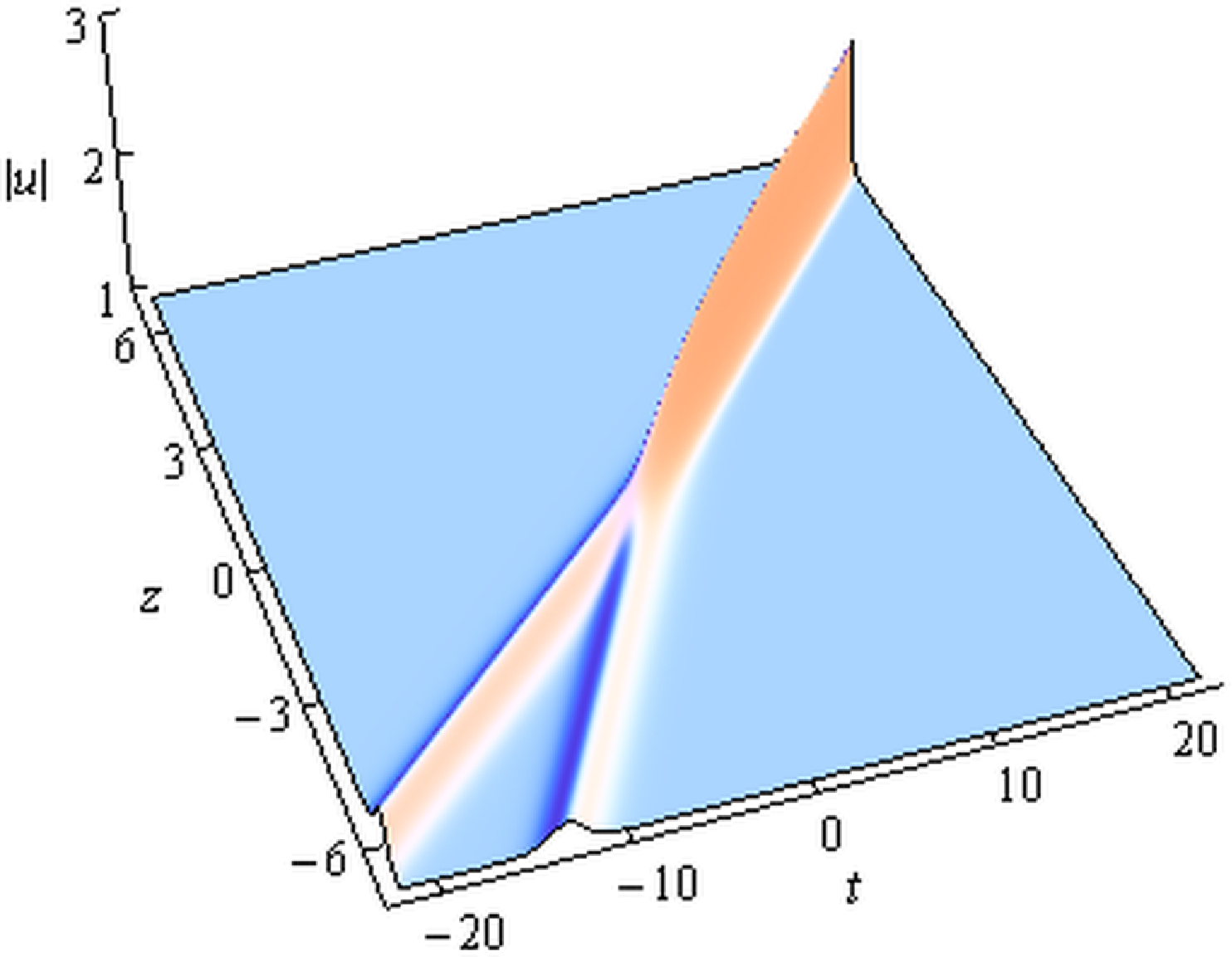}}\hfill
\subfigure[]{ \label{Fig4}
 \includegraphics[scale=0.45]{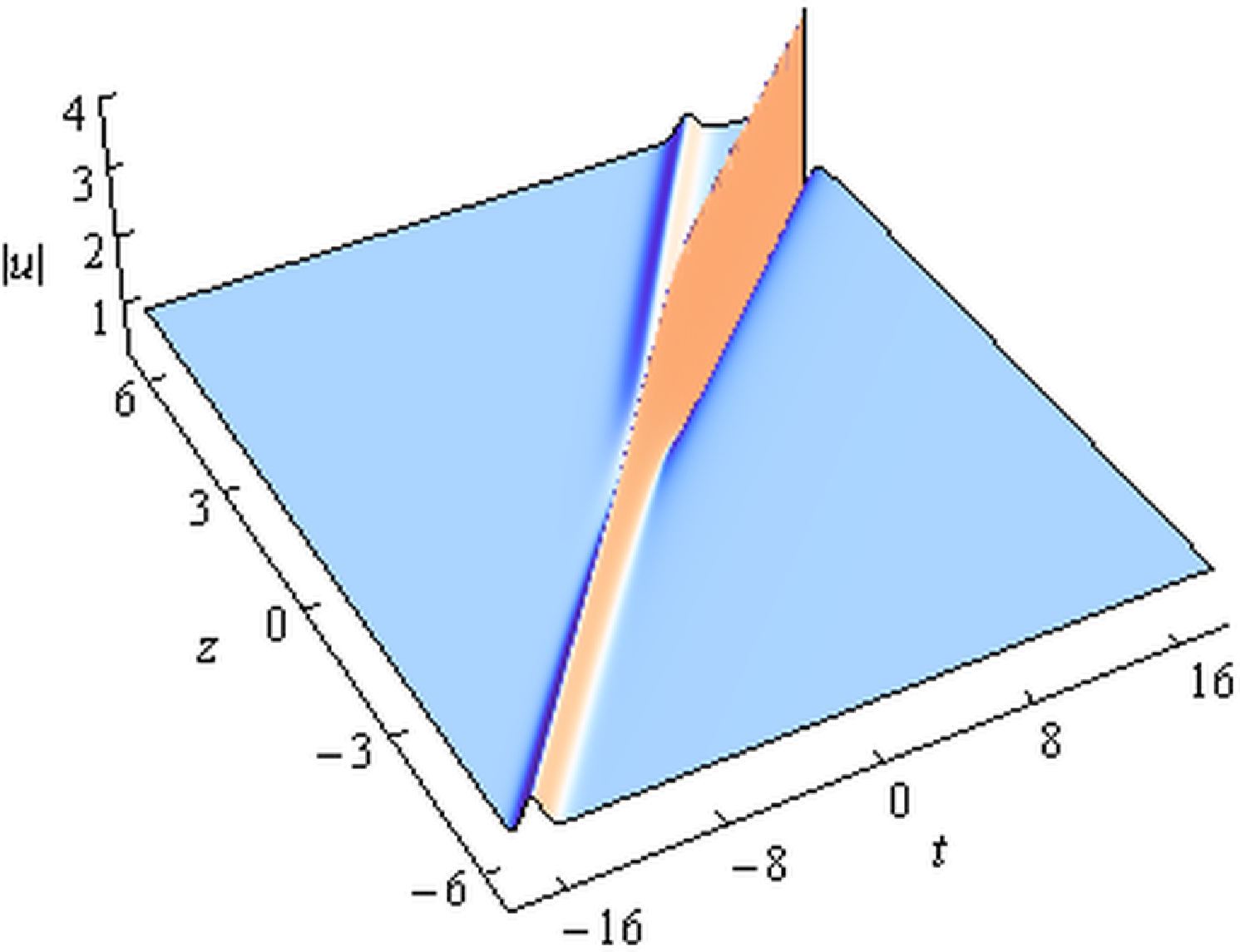}}
\caption{(a) Resonant $(2,1)$-interaction of three AD solitons with
$\lam_1=1.6$ and $\gamma_1=1+2\,\mi$. (b) Resonant
$(1,2)$-interaction of two AD solitons and one MH soliton with
$\lam_1=-1.42$ and $\gamma_1=1-2\,\mi$.  The other parameters are
chosen as $\alpha_1=1$, $\beta_1= 1+\mi$,  $\rho=1$, $\phi=0$ and
$\varepsilon=0.25$.}
\end{figure}

\emph{Resonant and elastic interactions.} --- If $\mu^{(i)}_1 \neq
0$ ($1\leq i\leq 3$) in solution~\eref{potentialTran2} with $N=1$,
the phase difference (which is neither $\pi$ nor $\pm
\frac{\pi}{2}$) between the embedded solution and the CW solution
results in that there are three asymptotic solitons appearing on top
of the same CW background. The asymptotic expressions of the three
solitons as $z\ra \pm \infty$ have the same form in
Eqs.~\eref{Solution1}--\eref{Solution3} except that
\begin{align}
& \phi^{(1)}_1= \frac{\pi}{2}\big[1 - \sgn
\big(\mu^{(2)}_1\mu^{(3)}_1 \big)\big], \quad \delta^{(1)}_1 =
\frac{1}{2}\,  \ln \left(\frac{|\mu^{(2)}_1|^2
\chi_1^-}{|\mu^{(3)}_1|^2\chi_1^+}\right), \\
& \phi^{(2)}_1= \frac{\pi}{2} \sgn\big(\mi \mu^{(1)}_1\mu^{(3)}_1
\big), \quad  \delta^{(2)}_1 = \frac{1}{2} \, \ln
\left(\frac{|\mu^{(1)}_1|^2\chi_1^2}{|\mu^{(3)}_1|^2 \lam_1\chi_1^+}\right),  \\
& \phi^{(3)}_1= -\frac{\pi}{2} \sgn\big(\mi \mu^{(1)}_1\mu^{(2)}_1
\big), \quad \delta^{(3)}_1 = \frac{1}{2}\,\ln \left(
\frac{|\mu^{(2)}_1|^2 \lam_1\chi_1^- }{|\mu^{(1)}_1|^2
\chi_1^2}\right).
\end{align}
Their wave numbers and frequencies can be respectively given as
follows: $(K^{(1)}_1, \Omega^{(1)}_1)=\big[-8 \varepsilon
\left(\rho^2+\lam_1^2\right)\chi _1-\frac{\chi _1}{6 \varepsilon },
2\chi_1\big]$, $(K^{(2)}_1,\Omega^{(2)}_1) =
\big[-\frac{\chi^+_1}{12 \varepsilon } -2\, \varepsilon (\chi_1^+)^2
\left(\lam_1+\chi_1^-\right), \lam_1+ \chi_1 \big]$, $(K^{(3)}_1,
\Omega^{(3)}_1)  =\big[\frac{\chi^-_1}{12 \varepsilon }+
2\,\varepsilon (\chi_1^-)^2 \left(\lam_1+\chi_1^+\right), \chi_1
-\lam_1\big] $, which  exactly satisfy the three-soliton resonant
conditions $K^{(1)}_1 =K^{(2)}_1 +K^{(3)}_1$ and $\Omega^{(1)}_1 =
\Omega^{(2)}_1 + \Omega^{(3)}_1 $. Associated with $\lam_1>0$ and
$\lam_1<0$, the solution can, respectively, exhibit the
$(2,1)$-resonant structure of two solitons merging into one soliton
and the $(1,2)$-resonant structure of one soliton diverging into two
solitons, as shown in Figs.~\ref{Fig3} and~\ref{Fig4}. When $\sgn
\big(\mu^{(2)}_1\mu^{(3)}_1 \big) =1 $, the three resonant solitons
all belongs to the AD case; while for $\sgn
\big(\mu^{(2)}_1\mu^{(3)}_1 \big) =-1 $, two are still the AD
solitons but the other one is of  the MH shape. That means that the
CW background exchanges its energy with one interacting soliton, and
causes such soliton changes its shape after resonant interaction, as
shown in Fig.~\ref{Fig4}.


\begin{figure}[H]
 \centering
\subfigure[]{\label{Fig5a}
\includegraphics[scale=0.45]{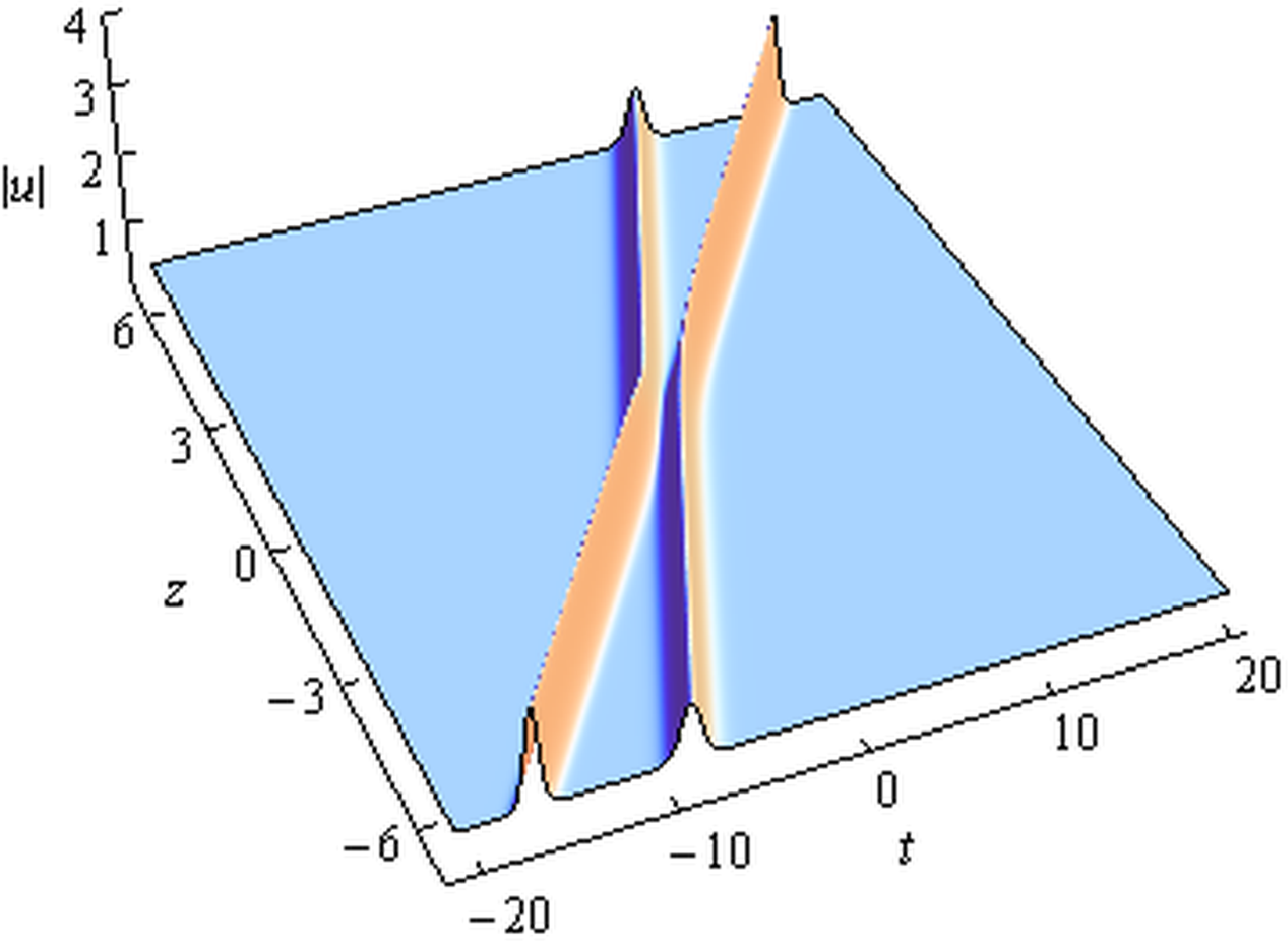}}\hfill
\subfigure[]{ \label{Fig5}
 \includegraphics[scale=0.45]{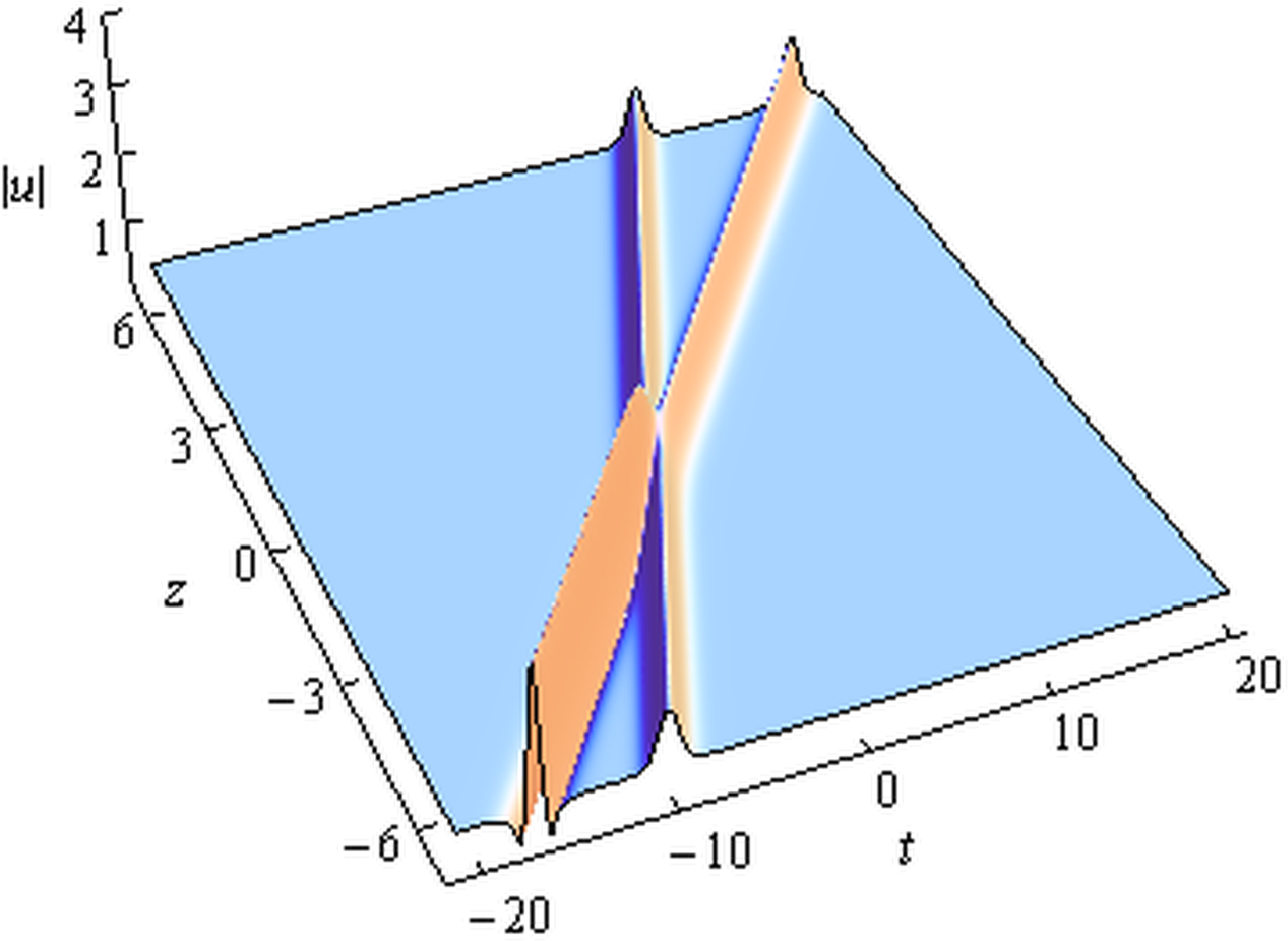}}
\caption{ (a) Shape-preserving elastic $(2,2)$-soliton interaction
with $\alpha_2=1+\mi$. (b) Shape-changing elastic $(2,2)$-soliton
interaction with $\alpha_2=1$.  The other parameters are chosen as
$\alpha_1=1$, $\beta_1= \mi$, $\beta_2=-1$, $\gamma_1=1$,
$\gamma_2=1- \mi$, $\rho=0.5$, $\lam_1=-1$, $\lam_2=1.42$, $\phi=0$
and $\varepsilon=0.25$. }
\end{figure}

On the other hand, one can also obtain the elastic $(2,2)$-soliton
interactions by implementing the DT two times. Our analysis shows
that there are four asymptotic solitons  as $z\ra \pm \infty$ in
solution~\eref{potentialTran2} with $N=2$ under the condition
$\mu^{(i)}_1 =\mu^{(j)}_2 =0 $ ($1\leq i,j \leq 3$).  The
expressions of asymptotic solitons are still of the form in
Eqs.~\eref{Solution1}--\eref{Solution3} except that there are some
difference for the parameters $\phi^{(i)}_k$ and $\delta^{(i)}_k$
($k=1,2 $; $1\leq i \leq 3$) (details are omitted for saving the
space). Each interacting soliton could be either the AD or MH one,
depending on the concrete parametric choice. For example,
Fig.~\ref{Fig5a} illustrates that the  AD solitons display the
standard elastic interaction, that is,  they can completely recover
their individual intensities and velocities after an interaction
except for the phase shift in their envelops. Note that the phase
shift, which corresponds to the instantaneous frequency at pulse
peak being nonzero, will result in the relative motion of
interacting solitons~\cite{Liu1}.  Also, the energy exchange may
take place between some interacting soliton and the CW background,
and result in the shape change of such soliton after interaction, as
seen in Fig.~\ref{Fig5}. However, this kind of soliton interactions
are still considered to be elastic in the sense that there is no
energy exchange between two different solitons.

\begin{figure}[H]
 \centering
\subfigure[]{     \label{Fig6}
\begin{overpic}[scale=0.45]{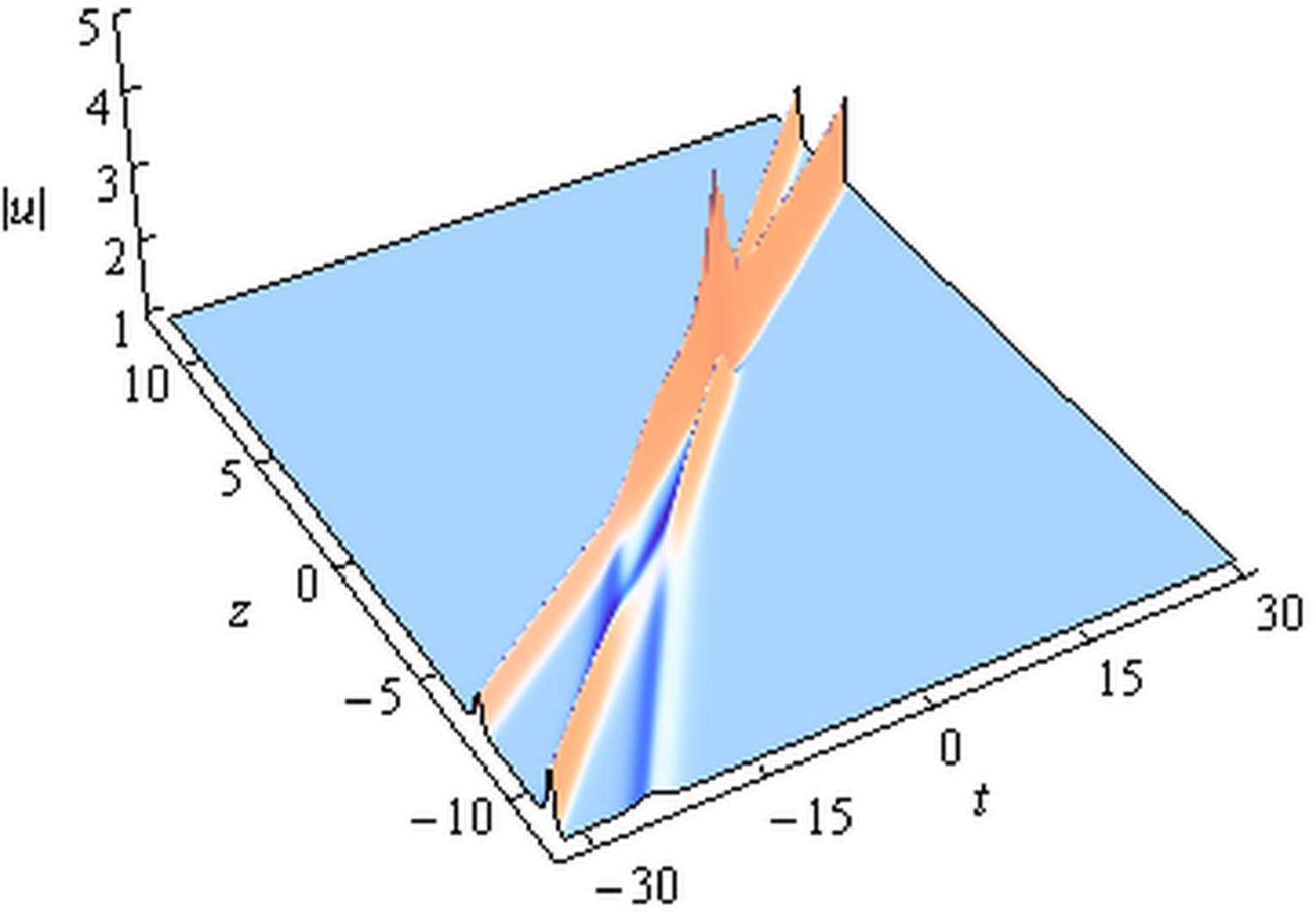}
\put(42,22){\color{red}\vector(0,-1){8}}
\put(60,72){\color{red}\vector(0,-1){8}}
\end{overpic} }\hfill
\subfigure[]{ \label{Fig7}
\begin{overpic}[scale=0.45]{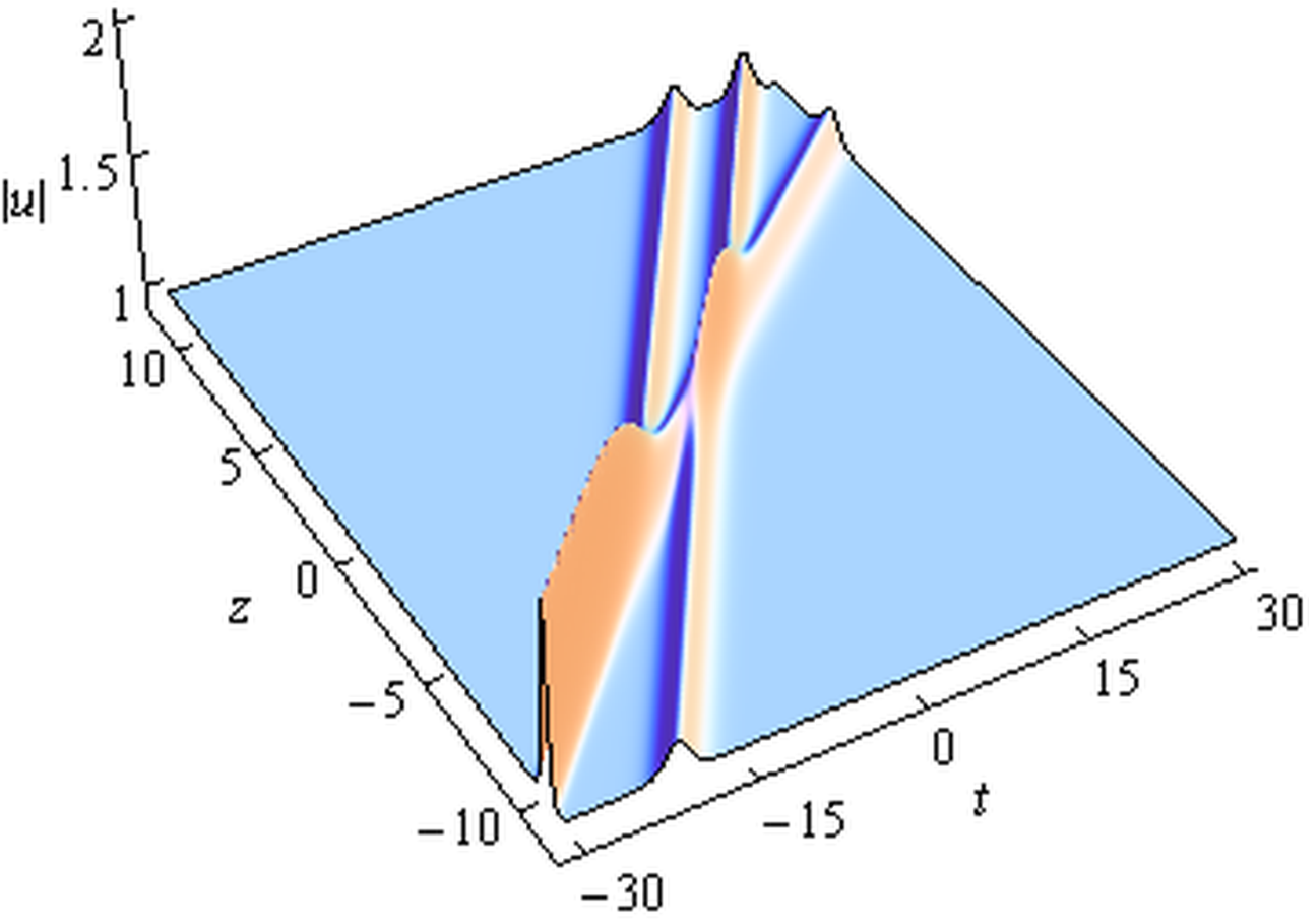}
\put(51.5,26){\color{red}\vector(0,-1){8}}
\put(57,76){\color{red}\vector(0,-1){8}}
\end{overpic} }
\caption{  (a) Inelastic $(3,2)$-soliton interaction with $\beta_1=
1+\mi$, $\beta_2=2+\mi$, $\gamma_1=0.2$, $\lam_1=1.5$ and
$\lam_2=-1.73$. (b) Inelastic $(2,3)$-soliton interaction with
$\beta_1= \mi$, $\beta_2=1+\mi$, $\gamma_1=1$, $\lam_1=1.73$ and
$\lam_2=-1.5$. The other parameters are chosen as
$\alpha_1=\alpha_2=\gamma_2=\rho=1$, $\phi=0$ and
$\varepsilon=0.25$. }
\end{figure}

\begin{figure}[H]
 \centering
\subfigure[]{\label{Fig8}
\includegraphics[scale=0.3]{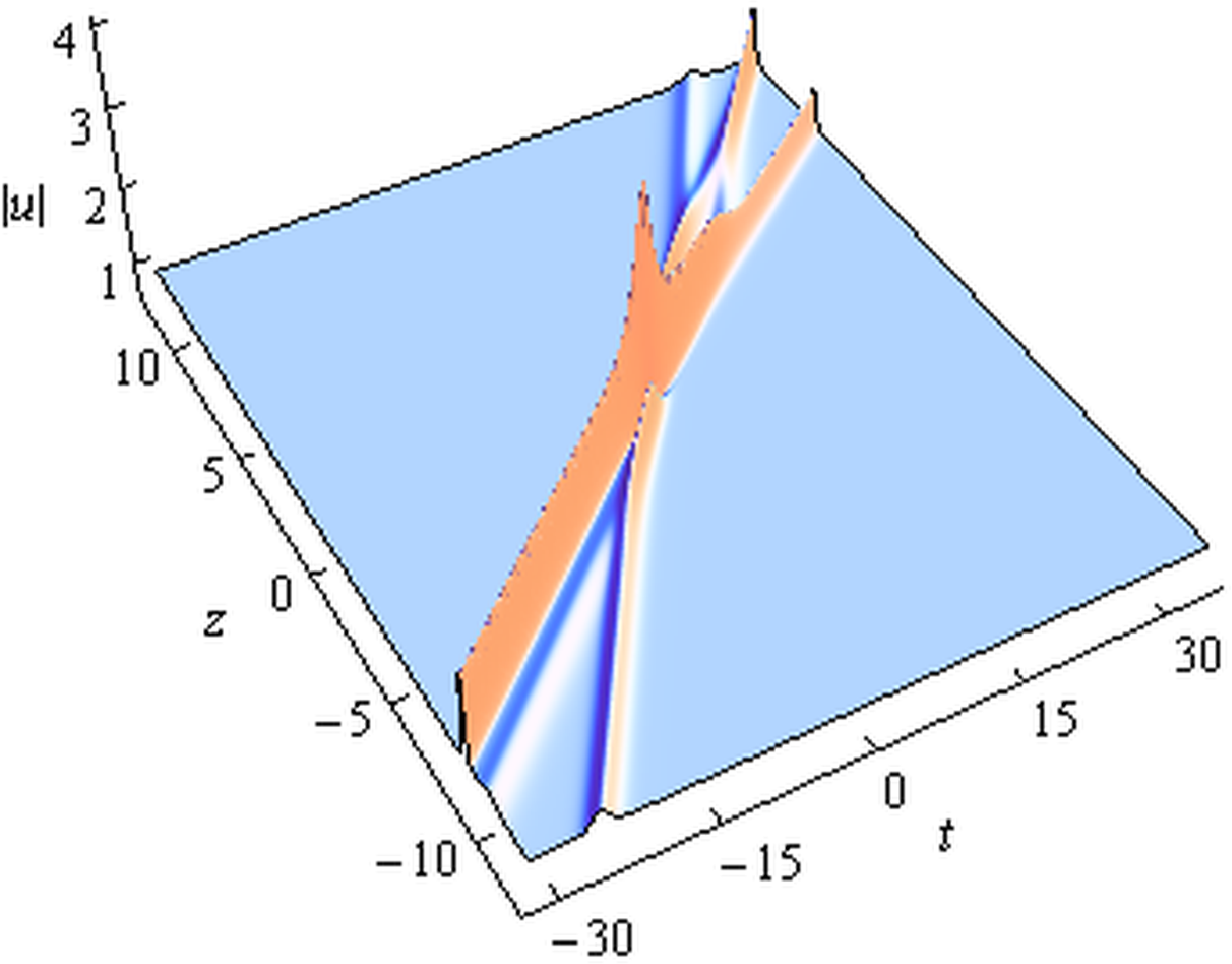}}\hfill
\subfigure[]{ \label{Fig9}
\includegraphics[scale=0.3]{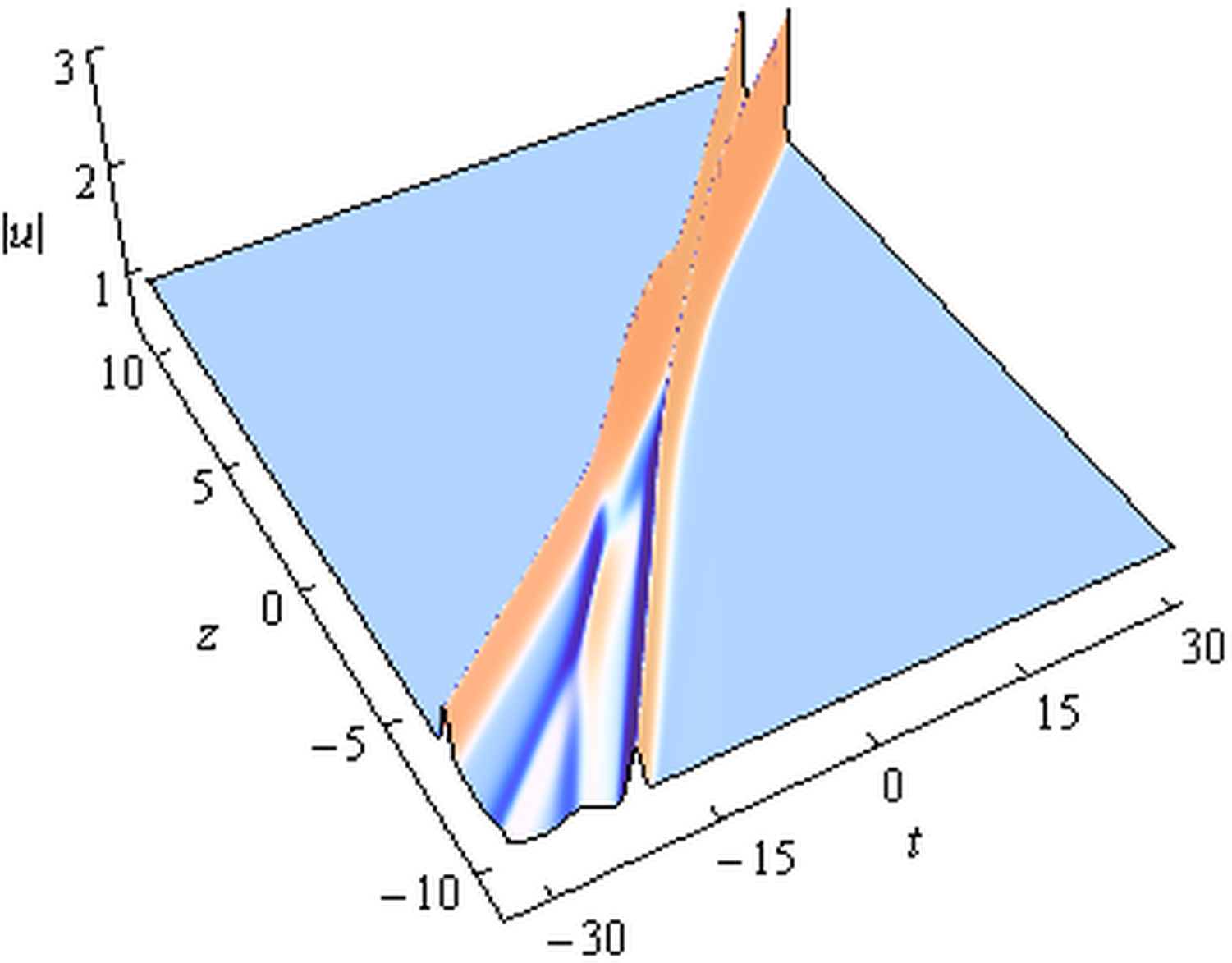}}\hfill
\subfigure[]{ \label{Fig10}
 \includegraphics[scale=0.3]{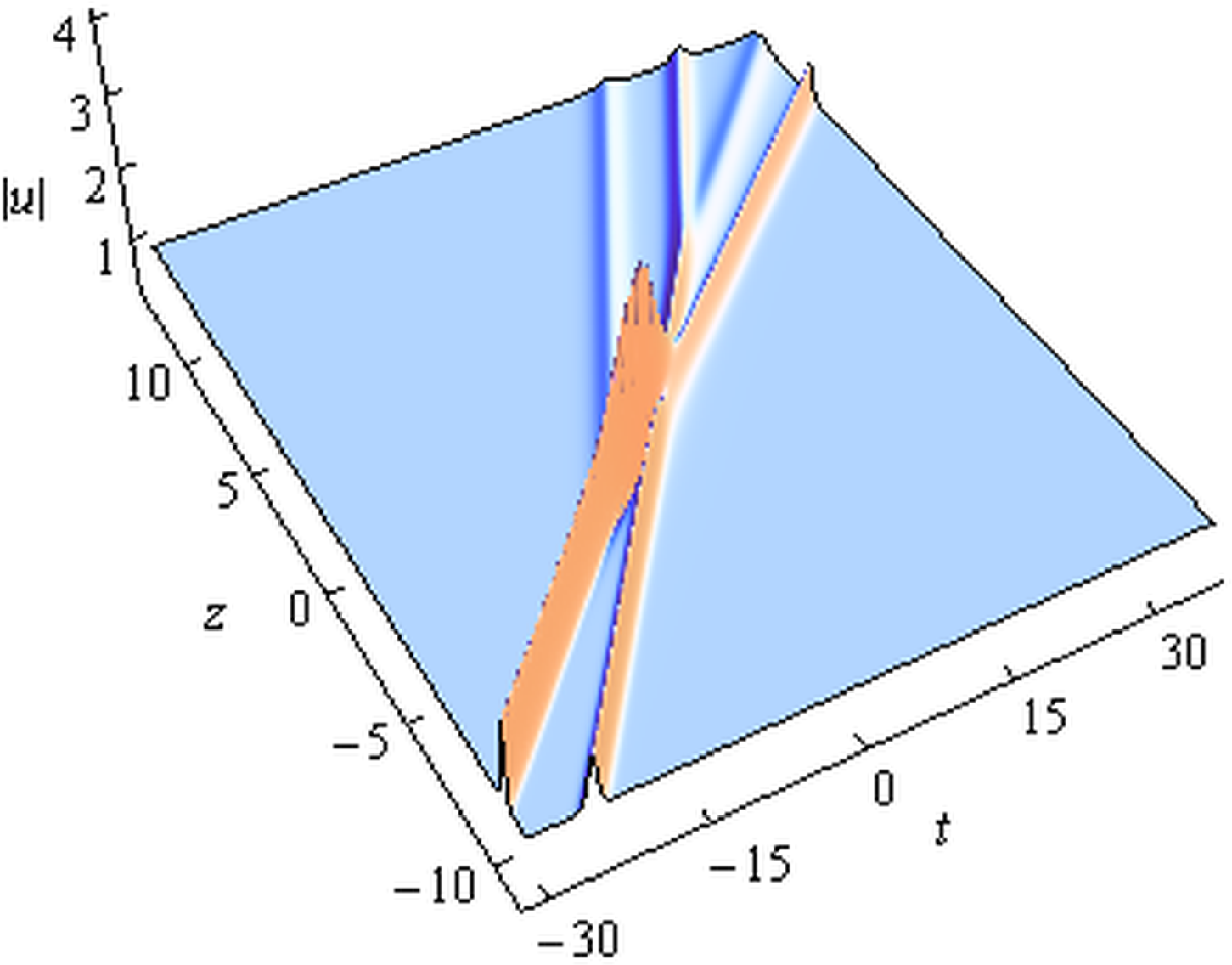}}
\caption{  (a) Inelastic $(3,3)$-soliton interaction with
$\alpha_2=4$, $\beta_1= -16\,\mi$, $\beta_2=1-5\,\mi$,
$\gamma_1=2-\mi$, $\gamma_2=0.5$, $\lam_1=1.5$ and $\lam_2=-1.73$.
(b) Inelastic $(4,2)$-soliton interaction with $\alpha_2=1-\mi$,
$\beta_1= 5\,\mi$, $\beta_2= 10\,\mi$, $\gamma_1=1+10\,\mi$,
$\gamma_2=-\mi$,  $\lam_1=1.5$ and $\lam_2=1.73$. (c) Inelastic
$(2,4)$-soliton interaction with  $\alpha_2=1-0.1\mi$,
$\beta_1=1-10\,\mi$,  $\beta_2= \mi$, $\gamma_1=1$, $\gamma_2= \mi$,
$\lam_1=-1.5$ and $\lam_2=-1.73$.  The other parameters are chosen
as $\alpha_1=1$, $\rho=1$, $\phi=0$ and $\varepsilon=0.25$.}
\end{figure}

\textit{Partially and completely inelastic interactions.}
--- For other cases in solution~\eref{potentialTran2} with $N=2$, one can obtain
five different types of inelastic soliton interactions. If there is
only one $\mu^{(i)}_k $ equal to $0$, the solution can exhibit the
$(3,2)$- and $(2,3)$-soliton interactions which are, respectively,
associated with $\lam_{k-3}>0$ and $\lam_{k-3}<0$ [see
Figs.~\ref{Fig6} and~\ref{Fig7}]. In both the two cases, the numbers
of interacting solitons as $z\ra \pm \infty$ are not equal, but one
$z\ra -\infty$ soliton and one  $z\ra \infty$ soliton [which are
marked by the red arrows in Figs.~\ref{Fig6} and~\ref{Fig7}] have
the same velocities and intensities and differ only by the phases of
their envelops. Accordingly, the $(3,2)$- and $(2,3)$-soliton
interactions belong to the \emph{partially inelastic type}. If none
of $\mu^{(i)}_k$'s ($1\leq i \leq 3$, $1\leq k\leq 2$) is equal to
$0$, the solution can display the $(3,3)$-, $(4,2)$- and
$(2,4)$-soliton interactions which are, respectively, associated
with $\lam_1\lam_2<0$, $\lam_1, \lam_2>0$ and $\lam_1,\lam_2<0$ [see
Figs.~\ref{Fig8}--\ref{Fig10}]. Such three interactions are of the
\emph{completely inelastic type} in the sense that the asymptotic
solitons as $z\ra -\infty$ totally differ from those as $z\ra
\infty$  in the velocities and intensities.

As for the above five inelastic  interactions, we find that in the
near-field region the asymptotic solitons connect with one another
via the ``X''- and ``Y''-type junctions, which correspond to the
elastic and resonant interactions, respectively. For example, there
are three resonant interactions and one elastic interaction in
Fig.~\ref{Fig6}, and there appear  three resonant interactions in
Fig.~\ref{Fig7}. Therefore, the ``X''-type elastic interaction and
``Y''-type resonant interaction are two fundamental structures to
form  various complicated soliton interactions in Eq.~\eref{SS},
where the numbers, velocities and intensities of interacting
solitons as $z\ra \pm \infty $ are in general not the same.

\emph{Conclusion.} ---
In this letter, via the DT method we have constructed new analytic
soliton solutions for Eq.~\eref{SS} which governs the propagation of
femtosecond  pulses in a monomode fiber with the TOD, SS and SRS
effects. We have revealed that two new types of femtosecond solitons
(i.e., the  AD and MH solitons) occur in Eq.~\eref{SS} with
$\sigma=1$ on a CW background. The numerical experiments have
indicated that the AD soliton can propagate stably for a long
distance with presence of a small initial perturbation or slight
violation of the fixed ratio of parameters in Eq.~\eref{SS}. More
importantly, we have obtained that the AD and MH solitons can
exhibit both the resonant and elastic interactions. Such two
fundamental interactions can generate various complicated
structures, in which the numbers, velocities and intensities of
interacting solitons are usually not the same before and after
interaction. In addition, we have found that some interacting
soliton may exchange its energy with the background in the
interaction, which results in one AD soliton changing into an MH
one, or one MH soliton into an AD one. It should be noted that
changing the propagation direction of optical solitons is an
important concept for realizing optical
switching~\cite{Opticalsolitons}. Therefore, as a self-induced
Y-junction waveguide, the soliton resonant interaction might bring
about some applications in all-optical information processing and
routing of optical signals~\cite{NonlOptics,Opticalsolitons}. In
mathematics, our results will enrich the knowledge of soliton
interactions in a (1+1)-dimensional integrable equation with the
single field. It is also worthy of being studied to make a finer
classification of soliton interactions in Eq.~\eref{SS} with $
\sigma=1 $.


\emph{Acknowledgements.} --- This work has been supported by the
Science Foundations of China University of Petroleum, Beijing (Grant
No. BJ-2011-04), by the National Natural Science Foundations of
China under Grant Nos. 11247267, 11371371, 11426105, 61475198,  and
by the Fundamental Research Funds of the Central Universities
(Project No. 2014QN30).

\end{document}